\documentclass{article}
\usepackage[utf8]{inputenc}
\usepackage{graphicx}
\usepackage{times}
\usepackage{xcolor}
\usepackage{amsmath}
\usepackage{subfigure}
\usepackage{amssymb}
\usepackage{epstopdf}
\usepackage[numbers,sort&compress]{natbib}
\usepackage{appendix}
\usepackage{multirow}

\newcommand{\bea}{\begin{eqnarray}}
\newcommand{\eea}{\end{eqnarray}}

\begin{document}

\title{Quantization of the Bateman damping system  with conformable derivative  }

\author{Tariq AlBanwa, Ahmed Al-Jamel, Eqab.M.Rabei and  Mohamed.Al-Masaeed\\
Physics Department, Faculty of Science, Al al-Bayt University,\\ P.O. Box 130040, Mafraq 25113, Jordan\\albanwatariq@gmail.com\\aaljamel@aabu.edu.jo, aaljamel@gmail.com\\eqabrabei@gmail.com\\
 moh.almssaeed@gmail.com}

\maketitle


\begin{abstract}
In this work, the conformable Bateman Lagrangian for the damped harmonic oscillator system is proposed using the conformable  derivative concept. In other words, the integer derivatives are replaced by conformable  derivatives of order $\alpha$ with $0<\alpha\leq 1$. The corresponding conformable Euler-Lagrange equations of motion and fractional Hamiltonian are then obtained. The system is then canonically quantized and the conformable Schrodinger equation is constructed. The fractional-order dependence of the energy eigenvalues $E_n ^\alpha$  and eigenfunctions $\psi_n ^\alpha$ are obtained using  using suitable transformations and the extended fractional Nikiforov-Uvarov method. The corresponding conformable continuity equation is also derived and the probability density and probability current are thus suitably defined. The probability density evolution as well as its dependence on $\alpha$ is plotted and analyzed for various situations. It is found that the energy eigenvalues are real and there are sort of gradual ordering in the behavior of the probability densities. 
\\

\end{abstract}

\section{Introduction}
Dissipation is an inescapable part in all real physical systems, from classical surfaces in relative motion to quantum systems such as molecules, atoms, nuclei and radiating point charge. Such non-conservative systems have been studied several decades ago by researchers 
. Historically,  Bateman \cite{batemanDissipativeSystemsRelated1931} suggested a Lagrangian for the damped harmonic oscillator that leads to the exact equations of motion. After that, the Hamiltonian corresponding to Bateman's Lagrangian is constructed independently by Cardirola and Kanai \cite{caldirolaForzeNonConservative1941,kanaiQuantizationDissipativeSystems1948}. Because of the explicit time-dependence Lagrangian and Hamiltonian of such systems, their quantization is not an easy task.  This topic on the quantization of non-conservative systems has attracted many researchers. In \cite{jainPathIntegralsWavepacket2007}, the Lagrangian for damped mechanical systems with various forms of dissipation are quantized using the path integral formalism. In \cite{baldiottiQuantizationDampedHarmonic2011}, they revisited the description of the damped harmonic oscillator with an assessment of previous works mainly based on the Bateman-Caldirola-Kanai model and a new model has been proposed that has better energy behavior and relate it to some existing open-systems approaches. In \cite{markusQuantumParticleMotion2016}, the Feynman path integral is applied and widened toward the calculation of the kernel of a quantum damped harmonic oscillator. Besides, in \cite{serhanQuantizationDampedHarmonic2018}, a suitable Hamiltonian that describes the damped harmonic oscillator is constructed starting from Bateman Lagrangian. Also, the Hamilton-Jacobi equation is written and the action function is obtained. Then, the system is quantized using the WKB approximation and the canonical quantization. In addition, K. Takahashi \cite{takahashiQuantizationMassiveBateman2018} performed quantization on the dual Bateman's system (BDS) by decomposing it into two effectively independent massless subsystems with reduced degrees
of freedom. The original massive BDS that satisfies the canonical quantization condition is then rebuilt by superposing the two massless subsystems.

In the last two decades, the quantization of physical systems is extended to the framework of fractional calculus by many researchers and has become of prime important in physics. The theory of fractional calculus is as old as classical calculus and classified as generalized
fractional integrals or derivatives. There are different definitions of fractional derivative that are proposed, and the most popular definitions are Riemann-Liouville, Riesz, and Caputo definitions. Each definition has some characteristic properties; for general review see \cite{oldham1974fractional,miller1993introduction,kilbas2006theory,klimek2002lagrangean,agrawal2002formulation}. Researchers are paying
attention to the implementation of fractional calculus as they found that fractional order derivatives are useful in the description of many physical phenomena in the real world. For instance, Riewe \cite{rieweNonconservativeLagrangianHamiltonian1996,rieweMechanicsFractionalDerivatives1997} , Rabei \textit{et al.} \cite{rabeiQuantizationFractionalSystems2010,rabeiQuantizationBrownianMotion2006} and many others used the fractional calculus techniques to construct
the Lagrangian and Hamiltonian for the non conservative systems. Rabei et.al \cite{rabei2009fractional} discussed how to find the solution of Schrodinger equation for some systems that have a fractional behavior in their Lagrangian and obey the WKB approximation. Besides, the canonical quantization of a system with Brownian motion is carried out using fractional calculus by Rabei et.al \cite{rabei2006quantization}. 

In 2014, Khalil et al. \cite{khalilNewDefinitionFractional2014} suggested a modern fashioned fractional derivative termed as
the conformable  derivative (CD), which is defined using the usual fundamental limit definition of the classical derivative rather than in terms of integrals as in the other definitions. Given a function $f\in [0,\infty) \to {R}$. The conformable  derivative of $f$ with order $\alpha$ is defined by
\begin{equation}
D^{\alpha}[f(t)]=\lim_{\epsilon\rightarrow 0}\frac{f(t+\epsilon t^{1-\alpha})-f(t)}{\epsilon},~~~~ t>0,
\end{equation}
with $0<\alpha\leq 1$. This CD operator is linear and satisfies the general properties of integer order derivatives, such as the formula of the derivative of the product or quotient of two functions and the chain rule, which makes it favorable over the traditional fractional derivatives \cite{khalilNewDefinitionFractional2014}. In this paper, we adopt $D^\alpha f$ to denote the conformable  derivative (CD) of $f$ of order $\alpha$. 

Many researchers considered this new definition as a more sophisticated candidate to the extension of the classical
derivative to the fractional domain. The CD has found various applications in the physical sciences. For instance, the heat conformable  differential equation is investigated and the exact solutions are searched in \cite{khalil2014conformable}. Also, the conformable  Euler-Lagrange equation and Hamiltonian formulation were discussed in \cite{lazo2016variational}. The deformation of the ordinary quantum mechanics using the concept of conformable  calculus is considered by Chung et.al \cite{chung2020effect}. The Authors defined two fundamental operators, namely, the $\alpha$-position operator and $\alpha$-momentum operator, and then the $\alpha$-Hamiltonian  operator is constructed and the related conformable Schrodinger equation is reached. They also presented a formulation for the conformable quantum mechanics boosted by some illustrative physical applications. In \cite{al2019search}, the search for fractional ordering in the mass spectra of heavy quarkonia is investigated with a conformable  derivative potential model. The saturation effects due to non-linearity effects is discussed in these short-lived bound states. The CD is used by \cite{karayerConformableFractionalNikiforov2016} to define the conformable  Schrodinger equation, and the conformable  Nikiforov-Uvarov method is introduced to obtain the energy eigenvalues and eigenfunctions. Also, it has been used in \cite{karayerSolutionsLocalFractional2019} to study the sine-Gordon equation to obtain exact solitary wave solutions within the frame of conformable calculus. In \cite{chungFractionalNewtonMechanics2015}, CFD is used to study the conformable Newtonian mechanics, where the conformable calculus of variations is introduced and the conformable Euler–Lagrange equation is constructed. The CD is  also used by \cite{rabei2018quantization} to study the fractional singular Lagrangian system. They obtained the equations of motion and determined the action integral after. The fractional Christ-Lee model is then discussed and quantized using WKB approximation to demonstrate the applicability of their work. Using conformable calculus, the approximation methods employed in quantum mechanics have recently been extended to become usable in conformable quantum mechanics (Variational method \cite{https://doi.org/10.1002/mma.7963} , Perturbation theory \cite{al2021extension} and WKB approximation \cite{doi:10.1142/S0217732322501449} ). In addition the conformable harmonic oscillator is quantized by using $\alpha$ -creation and $\alpha$ -annihilation operators \cite{al2021quantization}. Furthermore, the deformation of special relativity is articulated in the context of conformable derivatives \cite{al-jamel_effect_2022}. Recently, more than one equation has been solved and the behavior of the solution is studied using conformable calculus such as (Laguerre differential equation \cite{https://doi.org/10.48550/arxiv.2112.01322}, Angular equation of the Schrodinger Equation \cite{https://doi.org/10.48550/arxiv.2203.11615} and Schrodinger equation with Hydrogen atom \cite{https://doi.org/10.48550/arxiv.2209.02699}  )

In this paper, we propose a conformable  Lagrangian as a natural extension to the classical Bateman Lagrangian \cite{batemanDissipativeSystemsRelated1931}. The corresponding equations of motions and conformable  Hamiltonian will then be obtained. The canonical quantization for conformable  systems as described in \cite{chung2020effect} will be used to obtain the possible  energy eiegnvalues and eigenfunctions in terms of the $\alpha$ order. In the sequel, we review in section \ref{sec1} the theoretical tools needed in our study. In section \ref{sec2}, formalism and quantization procedure will be presented. Final in section \ref{sec3} a summary and conclusions will be given.
\section{The extended Nikiforov-Uvarov method with conformable  derivative (ENU-CD)}
\label{sec1}
The extended Nikiforov-Uvarov method (ENU) is a generalization of the Nikiforov-Uvarov method  to obtain the eigenvalues and eigenfunctions of differential equations that can be transformed into hypergeometric form \cite{karayerExtensionNikiforovUvarovMethod2015}. It has been used in some research articles in quantum mechanics to obtain the eigenvalues and eigenfunctions of the wave equation \cite{al-jamelSaturationHeavyQuarkonia2018b,karayerSolutionSchrodingerEquation2018}. Consider the conformable  differential equation of the standard form \cite{karayerConformableFractionalNikiforov2016,karayerExtensionNikiforovUvarovMethod2015}: 
\begin{equation}
\label{eq1} 
D^{\alpha}[D^{\alpha}\psi(s)]+\frac{\tilde{\tau}(s)}{\sigma }D^{\alpha}\psi (s)+\frac{\tilde{\sigma}(s)}{\sigma ^{2}(s)}\psi(s)=0
\end{equation}
where $\tilde{\tau}(s)$, $\sigma(s)$ and $\tilde{\sigma}(s)$ are polynomials, at most second, third, and fourth degrees, respectively, then it can be solved analytically. Using the key property of CFD \cite{khalilNewDefinitionFractional2014,karayerExtensionNikiforovUvarovMethod2015}:
\begin{equation}
\label{D1}
D^{\alpha}{\psi}(s)=s^{1-\alpha}{\psi^{\prime}}(s)
\end{equation}
and
\begin{equation}
\label{D2}
D^{\alpha}[D^{\alpha}{\psi}(s)]=(1-\alpha)s^{1-2\alpha}\psi^{\prime}(s)+s^{2-2\alpha}\psi^{\prime\prime}(s).
\end{equation}
then Eq.(\ref{eq1}) turned into
\begin{equation}
\label{eq1b} 
\psi^{\prime\prime}(s)+\frac{(1-\alpha)\sigma (s) s^{-\alpha}+\tilde{\tau}(s)}{s^{1-\alpha}\sigma (s) }\psi^{\prime} (s)+\frac{\tilde{\sigma}(s)}{s^{2-2\alpha}\sigma ^{2}(s)}\psi(s)=0.
\end{equation}
Introducing the conformable  parameters $\tilde{\tau_f}(s)=(1-\alpha)\sigma (s) s^{-\alpha}+\tilde{\tau}(s)$, $\sigma_f (s)=s^{2-2\alpha}\sigma ^{2}(s) $, and $\tilde{\sigma_f}(s)=\sigma(s)$, then we obtain the standard form of equation of the conformable  ENU \cite{al-jamelSaturationHeavyQuarkonia2018b}
\begin{equation}
\label{eq2} 
\psi^{\prime\prime}(s)+\frac{\tilde{\tau_f}(s)}{\sigma_f (s) }\psi^{\prime} (s)+\frac{\tilde{\sigma}(s)}{\sigma_f ^{2}(s)}\psi(s)=0
\end{equation}
The next natural step is to to propose the Ansatz
\begin{equation}
\label{eq2b}
\psi(s)=\phi(s)Y(s),
\end{equation}
which reduces Eq.(\ref{eq2}) to the hypergeometric form,
\begin{equation}
\label{eq3} 
\sigma_f(s)Y^{\prime \prime }+\tau(s)Y^{\prime }+h(s) Y(s)=0.
\end{equation}
Here $\phi(s)$ fulfills the first order differential equation
\begin{equation}
\label{eq4}
\frac{\phi^{\prime } (s)}{\phi (s)}=\frac{\pi_f(s)}{\sigma_f (s)},
\end{equation}
and $h(s)$ fulfills
\begin{equation}
\label{eq4b}
h(s)= \pi_f^{\prime}(s)+G(s).
\end{equation}
Here $Y(s)$ is a type of hypergeometric functions whose polynomial solutions satisfy a Rodriguez formula of the form
\begin{equation}
\label{eq5}
Y_{n}(s)=\frac{B_{n}}{\rho (s)}\frac{d^{n}}{ds^{n}}\left[ \sigma_f ^{n}(s)\rho (s)\right],
\end{equation}
where $B_{n}$ is the normalization constant, and $\rho$ is called the weight or density function and must satisfy the condition
\begin{equation}
\label{eq6}
 (\sigma_f \rho )^{\prime}=\tau \rho.
\end{equation}
The function $\pi_f(s)$ and the function $G(s)$ required for this method are defined through the following relation
\begin{equation}
\label{eq7}
\pi_f =\frac{\sigma_f ^{\prime }(s)-\tilde{\tau_f} (s)
}{2}\pm
\sqrt{\left(\frac{\sigma_f^{\prime}(s)-\tilde{\tau_f}(s)}{2}\right)^{2}-\tilde{\sigma_f}(s)+G(s){\sigma_f(s)}}, 
\end{equation}
and are chosen so that the function $\pi_f(s)$ is a polynomial of at most $2\alpha$ degree. Also, the function $h_n(s)$ is determined from the relation
\begin{equation}
\label{eq8b}
h_n(s) =-\frac{n}{2}\tau ^{\prime
}(s)-\frac{n(n-1)}{6}\sigma_f ^{\prime \prime}(s)+C_n,~~~(n=0,1,2,...)
\end{equation}
where
\begin{equation}
\label{eq9}
\tau (s)=\tilde{\tau_f }(s)+2\pi_f (s).
\end{equation}
Then, the equality of Eq.(\ref{eq4b}) and Eq.(\ref{eq8b}) leads to the energy eigenvalues.
\section{Formalism}
\label{sec2}
Consider Bateman's Lagrangian \cite{batemanDissipativeSystemsRelated1931}
\begin{equation}
L=\frac{m}{2}\left(\dot{q}^{2}-\omega ^{2}q^{2}\right)e^{\lambda t}.
\end{equation}
This Lagrangian describes the one dimensional damped harmonic oscillator. We define the corresponding conformable  Bateman Lagrangian as
\begin{equation}
\label{fBL}
     L(q^\alpha,D^{\alpha}_t q^\alpha, t^{\alpha})=\frac{m^\alpha}{2}  ([D^{\alpha}_t q^\alpha]^2-\omega^{2\alpha} q^{2\alpha}) e^\frac{\lambda t^ \alpha}{\alpha}  
\end{equation}

Introduce a new coordinate $y^\alpha$ by
\begin{equation}
     y^\alpha= q^\alpha e^\frac{\lambda t^ \alpha}{2\alpha}  
   \end{equation} 
	Then,
\begin{equation}
    q^\alpha= \frac{y^\alpha }{e^\frac{\lambda t^ \alpha}{\alpha}}
\end{equation}
Operating on both sides by $D^\alpha_t$, we have
\begin{equation}
    D^{\alpha}_t q^\alpha =\frac{e^\frac{\lambda t^ \alpha}{2\alpha}(D^\alpha_t y^\alpha - y^\alpha \frac{\lambda}{2})}{e^\frac{\lambda t^ \alpha}{\alpha}}
\end{equation}
Substituting this result in Eq.(\ref{fBL}), and do some little algebra, we obtain
\begin{equation}
     L(y^\alpha,D^{\alpha}_t y^\alpha)= \frac{m^\alpha}{2} ([D^\alpha_t y^\alpha]^2 +[\frac{\lambda^2}{4}-\omega^{2\alpha}]y^{2\alpha}-y^{\alpha}\lambda D^{\alpha}_t y^\alpha ).
\end{equation}
The conformable  Euler-Lagrange equation of motion can be obtained using 
\begin{equation}
   \frac{\partial L}{\partial y^{\alpha}} - D^{\alpha}_t (\frac{\partial L}{\partial [D^{\alpha}_t y^{\alpha}])}= 0.
\end{equation} 
By noting that
\begin{eqnarray}
    \frac{\partial L}{\partial y^{\alpha}} &=& m^\alpha[\frac{\lambda^2}{4}-\omega^{2\alpha}]y^{\alpha}-\frac{m^\alpha \lambda D^\alpha_t y^\alpha}{2} \\\nonumber
  \frac{\partial L}{\partial \left[D^{\alpha}_t y^{\alpha}\right]} &=& m^{\alpha}D^{\alpha}_t y^{\alpha} -\frac{m^{\alpha} \lambda y^{\alpha}}{2}  \\\nonumber
 D^{\alpha}_t \frac{\partial L}{\partial [D^{\alpha}_t y^{\alpha}]} &=& m^{\alpha} D^{\alpha}_t D^{\alpha}_t y^{\alpha}- \frac{m^{\alpha} \lambda D^{\alpha} y^{\alpha}}{2}\nonumber
\end{eqnarray} 
	The equation of motion reads as
 \begin{equation}
    D^{\alpha}_t D^{\alpha}_t y^{\alpha}+(\frac{\lambda^2}{4}-\omega^{2\alpha})y^{\alpha}=0
\end{equation}
The Hamiltonian is defined as
\begin{equation}
    H(y^\alpha,P^\alpha_y) =P^\alpha_y D^{\alpha}_t y^\alpha  -L. 
\end{equation}
The momentum operator is defined by \cite{mozaffari2018conformable}
\begin{equation}
    P^{\alpha}_y = \frac{\partial L}{\partial [D^{\alpha}_t y^{\alpha}]}= m^{\alpha}  D^{\alpha}_t y^{\alpha}- \frac{m^{\alpha} \lambda  y^{\alpha}}{2}
\end{equation}
Then, the Hamiltonian becomes
\begin{equation}
\label{H}
     H(y^{\alpha},P^{\alpha}_y)= \frac{(P^\alpha_y )^2}{2m^\alpha}+ \frac{1}{2} m^\alpha \omega^{2\alpha} y^{2\alpha}+ \frac{1}{2}\lambda y^\alpha P^\alpha_y.
\end{equation}

\begin{figure}[!htb]
		 	 	\centering
		 	 	\subfigure[]{
		 	 		\includegraphics[scale=0.85]{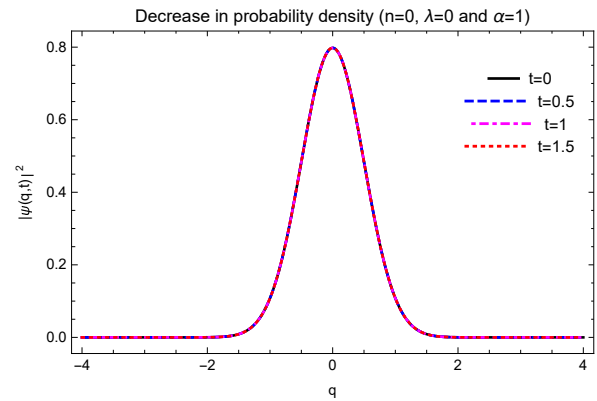}
		 	 	}
		 	 	\subfigure[]{
		 	 		\includegraphics[scale=0.85]{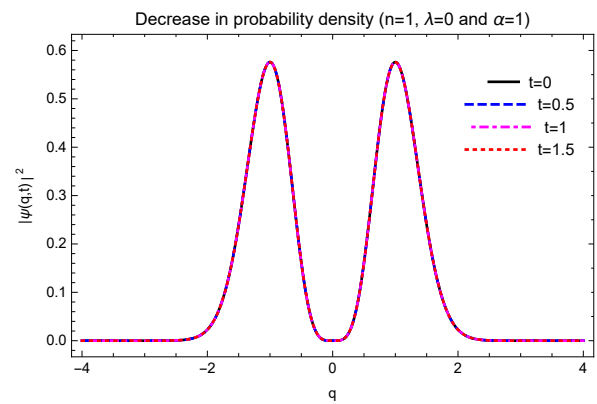}
		 	 	}
\caption{The probability density time evolution for the ground state $n=0$  and the first excited state $n=1$  with parameter set $\lambda=0$ and $\alpha=1$.}
\label{fig1}
 \end{figure}
\begin{figure}[!htb]
		 	 	\centering
		 	 	\subfigure[]{
		 	 		\includegraphics[scale=0.85]{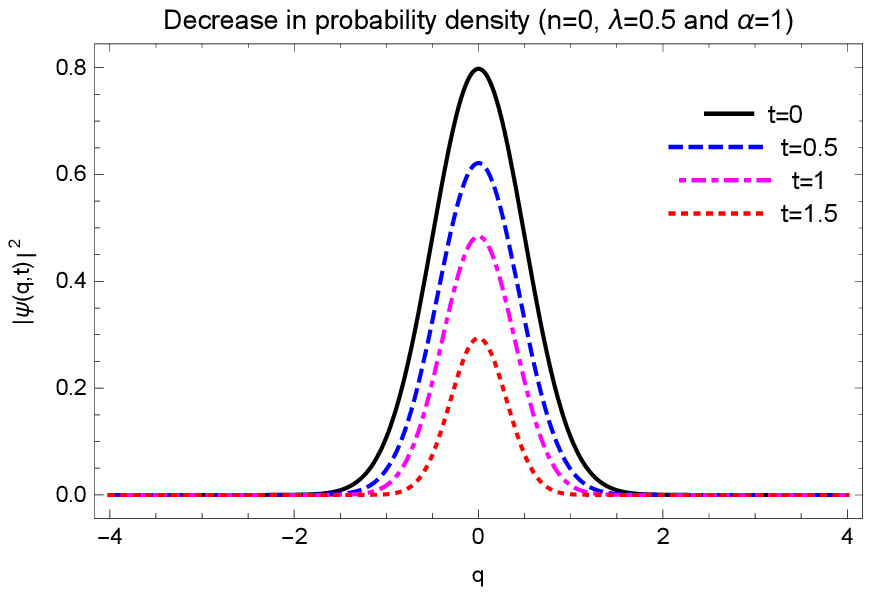}
		 	 	}
		 	 	\subfigure[]{
		 	 		\includegraphics[scale=0.85]{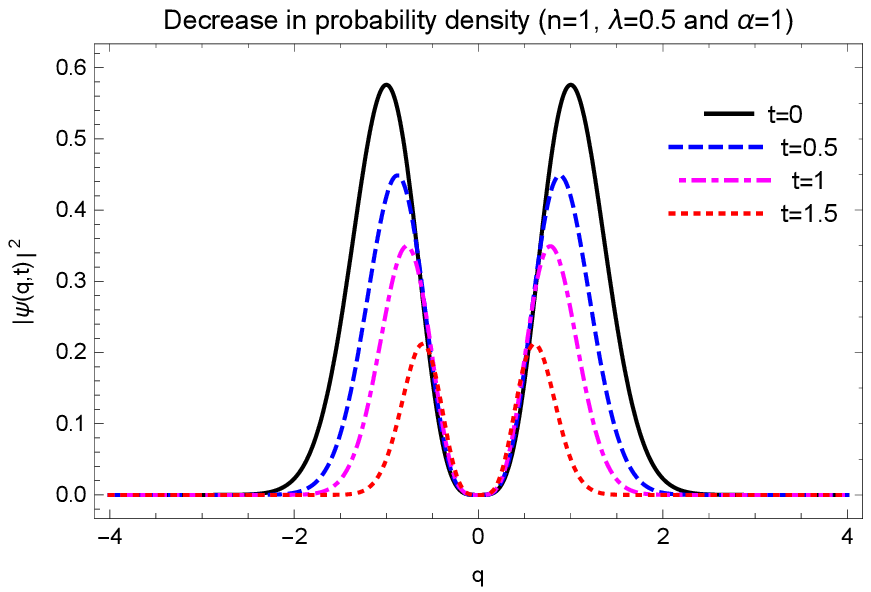}
		 	 	}
\caption{The probability density time evolution for the ground state $n=0$  and the first excited state $n=1$ with parameter set $\lambda=0.5$ and $\alpha=1$.}
\label{fig2}
 \end{figure}
\begin{figure}[!htb]
		 	 	\centering
		 	 	\subfigure[]{
		 	 		\includegraphics[scale=0.75]{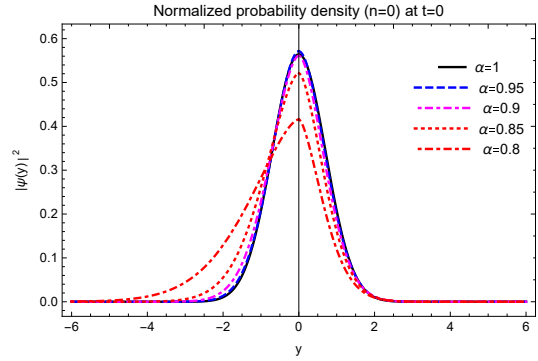}
		 	 	}
		 	 	\subfigure[]{
		 	 		\includegraphics[scale=0.75]{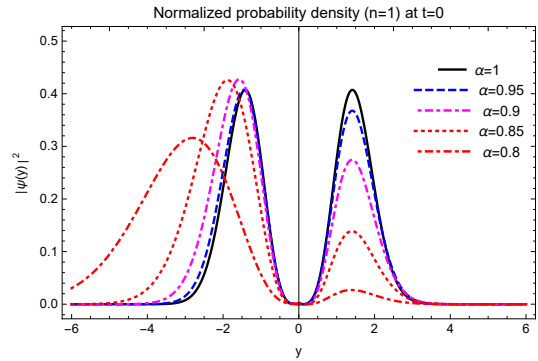}
		 	 	}
				\subfigure[]{
		 	 		\includegraphics[scale=0.75]{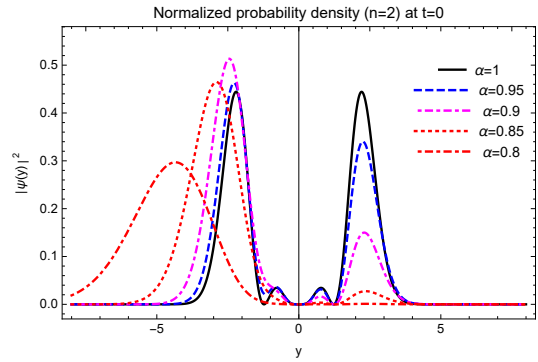}
		 	 	}
				\subfigure[]{
		 	 		\includegraphics[scale=0.75]{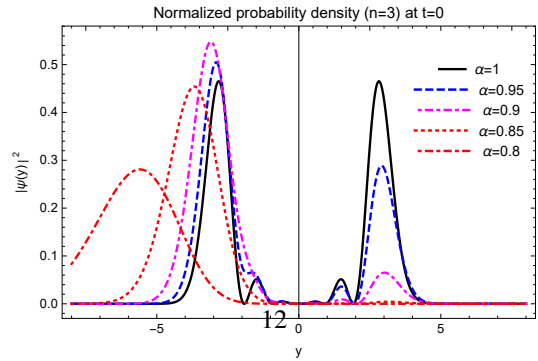}
		 	 	}
		 	 	
\caption{The behavior of the normalized probability density at time $t=0$ in terms of the fractional order $\alpha$ for the states $n=0,1,2,3$. The parameter values are chosen such that $F=\frac{m^\alpha}{\hbar^\alpha}\sqrt{\omega^{2\alpha} - \frac{\lambda^2}{4}}=1$.}
\label{fig3}
 \end{figure}

To apply canonical quantization on the Hamiltonian Eq.(\ref{H}), we first notice that 
\begin{eqnarray}
    [P^\alpha_y+\frac{m^\alpha \lambda y^\alpha}{2}, e^{-f(y)}]\psi(y) &=& [P^\alpha_y,e^{-f(y)}]\psi(y) \\\nonumber
    &=&-i\hbar^\alpha D^\alpha_t[e^{-f(y)},\psi(y)]+i\hbar^\alpha  
    e^{-f(y)} D^\alpha_t\psi(y) \\\nonumber
    &=&-i\hbar^\alpha \psi(y) D^\alpha_t e^{-f(y)}
\end{eqnarray}
And using
\begin{equation}
    D^\alpha_t e^{-f(y)}= - y^{1-\alpha} f'(y)e^{-f(y)} 
\end{equation}
we obtain the commutator
\begin{equation}
    [P^\alpha_y+\frac{m^\alpha \lambda y^\alpha}{2}, e^{-f(y)}] =i\hbar^\alpha  y^{1-\alpha} f'(y)e^{-f(y)}. 
\end{equation}
Expanding the LHS of this commutator
\begin{equation}
    (P^\alpha_y + \frac{m^\alpha \lambda y^\alpha}{2})e^{-f(y)} - e^{-f(y)}(P^\alpha_y + \frac{m^\alpha \lambda y^\alpha}{2})=i\hbar^\alpha  y^{1-\alpha} f'(y)e^{-f(y)} 
\end{equation}
Then multiplying both sides from the left by $e^{+f(y)}$, and with a little algebra, yields
\begin{equation}
    e^{+f(y)}(P^\alpha_y + \frac{m^\alpha \lambda y^\alpha}{2})e^{-f(y)} = P^\alpha_y + \frac{m^\alpha \lambda y^\alpha}{2}+i\hbar^\alpha  y^{1-\alpha} f'(y)
\end{equation}
Choose $e^{+f(y)}$ such as
\begin{equation}
    f'(y)=-\frac{m^\alpha \lambda y^\alpha}{2}* \frac{1}{i\hbar^\alpha  y^{1-\alpha}}
\end{equation}
will give
\begin{equation}
    f(y)=\frac{i m^\alpha \lambda y^{2\alpha}}{4 \alpha \hbar^\alpha}
\end{equation}
Then
\begin{equation}
    e^\frac{+i m^\alpha \lambda y^{2\alpha}}{4 \alpha \hbar^\alpha}(P^\alpha_y + \frac{m^\alpha \lambda y^\alpha}{2}) e^\frac{-i m^\alpha \lambda y^{2\alpha}}{4 \alpha \hbar^\alpha} = P^\alpha_y
\end{equation}
Writing
\begin{equation}
    \eta= e^\frac{i m^\alpha \lambda y^{2\alpha}}{4 \alpha \hbar^\alpha}
\end{equation}
 we obtain the gauge transformation
\begin{equation}
 \eta (P^\alpha_y + \frac{m^\alpha \lambda y^\alpha}{2}) \eta^{-1} = P^\alpha_y.   
\end{equation}
This result can be generalized to
\begin{equation}
    \eta (P^\alpha_y + \frac{m^\alpha \lambda y^\alpha}{2})^2 \eta^{-1} =( P^\alpha_y)^2.
\end{equation}
Applying this on the Hamiltonian Eq.(\ref{H}), we have
\begin{equation}
\label{Hnew}
    \eta H \eta^{-1} \eta = \frac{(P^\alpha_y)^2}{2m^\alpha}+\frac{1}{2} m^\alpha (\omega^{2\alpha}-\frac{\lambda^2}{4}) y^{2\alpha}
\end{equation}
Thus, rather than applying canonical quantization on the Hamiltonian as given by Eq.(\ref{H}), we do this with the new Hamiltonian as given by the RHS of Eq.(\ref{Hnew}):
\begin{equation}
\label{Hnew2}
    H = \frac{(P^\alpha_y)^2}{2m^\alpha}+\frac{1}{2} m^\alpha (\omega^{2\alpha}-\frac{\lambda^2}{4}) y^{2\alpha}.
\end{equation}
Then, the corresponding conformable  Schrodinger equation is
\begin{equation}
   [ \frac{(P^\alpha_y)^2}{2m^\alpha}+\frac{1}{2} m^\alpha \Omega^2  y^{2\alpha}] \psi = E^\alpha  \psi
\end{equation}
where $\Omega^2 = \omega^{2\alpha} - \frac{\lambda^2}{4}$. On substituting for the conformable  momentum operator $P^\alpha_y =-i\hbar D^\alpha_y$, we find
\begin{equation}
   \frac{-\hbar^{2\alpha}}{2m^\alpha} (D^\alpha_y)^2 \psi+\frac{1}{2} m^\alpha \Omega^2  y^{2\alpha} \psi= E^\alpha  \psi
\end{equation}
Or

\begin{equation}
    (D^\alpha_y)^2 \psi= y^{2-2\alpha} \psi''+(1+\alpha) y^{1-2\alpha} \psi'
\end{equation}
Using Eq.(\ref{H}), we find that
\begin{equation}
\label{shr22}
    \psi''+\frac{(1+\alpha)}{y}\psi'+\frac{2m^\alpha}{y^{2} \hbar^{2\alpha} }(E^\alpha y^{2\alpha}- \frac{m^{\alpha} \Omega^2 }{2} y^{4\alpha})\psi = 0
\end{equation}
Then, we note that $\tau_f(y)=(1+\alpha)$, $\sigma_f=y$, and $\tilde{\sigma}_f(y)=\frac{2m^\alpha}{ \hbar^{2\alpha} }(E^\alpha y^{2\alpha}- \frac{m^{\alpha} \Omega^2 }{2} y^{4\alpha})$. Choosing $G(y)= S+P y^{\alpha-1}+Q y^{2\alpha-1}$, then one can show that $\pi_f$ takes the form:
\begin{equation}
    \pi_f(y)=\frac{\alpha}{2}+-\sqrt{(A+B y^{\alpha}+ F y^{2\alpha})^2}
\end{equation}
with  $A=\pm\frac{\alpha}{2}$, $S=0$, $B=0$, $P=0$, and $Q=\frac{2m^\alpha E^\alpha}{\hbar^{2\alpha}} y^{2\alpha}+B^2+2AF$, where $ F=\frac{m^\alpha}{\hbar^\alpha}\sqrt{\omega^{2\alpha} - \frac{\lambda^2}{4}}$. Then, using Eqs.(\ref{eq9}),(\ref{eq4b}) and (\ref{eq8b}), we obtain:
\begin{eqnarray}
\label{A}
    \tau(y)&=& 1+-2(A+F y^{2\alpha}) \\
		\label{B}
		 h_n(y)&=& -\frac{n}{2} \tau'- \frac{n(n-1)}{6}\sigma''_f +C_n \\
		\label{C}
		 h(y)&=&2n\alpha Fy^{2\alpha-1}+C_n. 		
\end{eqnarray}	
Equating Eq.(\ref{B}) with	Eq.(\ref{C}), we obtain 	
\begin{equation}
    E^\alpha = \alpha \hbar^\alpha \sqrt{\omega^{2\alpha}-\frac{\lambda^2}{4}} (n+\frac{1}{2} ),~~~n=0,1,2,3,...
\end{equation}
This result shows that the energy eigenvalues are real. Setting $\alpha=1$, the traditional damped harmonic oscillator energy eigenvalues are recovered, and in agreement with the result obtained in \cite{serhanQuantizationDampedHarmonic2018}. 
This result coincides with that found in \cite{alextension,al2021wkb,al2021extension} using different approach. The eigenfunctions can be found by finding firstly the needed functions from Eqs.(\ref{eq2b}-\ref{eq5}). The results are 
\begin{eqnarray}
    \phi(y) &=& y^{\alpha} e^{\frac{F}{2\alpha}y^{2\alpha}}  \\
		\rho(y) &=& y^{-\alpha} e^{\frac{F}{\alpha} y^{2\alpha}} \\
	  Y_n(y) &=& B_n y^{\alpha} e^{\frac{F}{\alpha}y^{2\alpha}}\frac{d^n}{d y^n}\left(y^{n-\alpha} e^{\frac{-F}{\alpha}y^{2\alpha}} \right)		
\end{eqnarray}			
Then, the eigenfunctions are 
\begin{equation}
\label{eq33}
    \psi_n(y)= B_n y^{\alpha} e^{\frac{F}{2\alpha}y^{2\alpha}} \frac{d^n}{d y^n}\left(y^{n-\alpha} e^{\frac{-F}{\alpha}y^{2\alpha}} \right),
\end{equation}
where $B_{n}$ is the normalization constant.
To define the probability current density and probability current correctly, we derive the continuity equation for Eq.(\ref{shr22}). Multiplying Eq.(\ref{shr22}) by $e^{\frac{-\lambda t^{\alpha }}{2}}$, and then following analogous procedure for deriving the continuity equation for traditional Schrodinger equation, we obtain 
\begin{equation}
D_{t}^{\alpha }\lbrack \psi ^{\ast}\psi e^{\frac{-\lambda t^{\alpha 
}}{2}}\rbrack +D_{y}^{\alpha }\lbrack \frac{\hbar^{\alpha }}{2im^{\alpha 
}}(\psi^{\ast}D_{y}^{\alpha }\psi -\psi D_{y}^{\alpha }\psi^{\ast}) 
e^{\frac{-\lambda t^{\alpha }}{2}}+\frac{\lambda }{2}y^{\alpha }\psi^{\ast}\psi e^{\frac{-\lambda t^{\alpha }}{2}}\rbrack =0.
\end{equation}
We define the probability density $\rho (y,t)$ and the probability 
current density $j(y,t)$ as
\begin{equation}
    \rho (y,t)=\psi^{\ast}\psi e^{\frac{-\lambda t^{\alpha }}{2}},
\end{equation}
and
\begin{equation}
j(y,t)=\frac{\hbar^{\alpha }}{2im^{\alpha 
}}(\psi^{\ast}D_{y}^{\alpha }\psi -\psi D_{y}^{\alpha }\psi^{\ast}) 
e^{\frac{-\lambda t^{\alpha }}{2}}+\frac{\lambda }{2}y^{\alpha }\psi^{\ast}\psi e^{\frac{-\lambda t^{\alpha }}{2}},
\end{equation}
respectively. Substituting from Eq.(\ref{eq33}), we obtain
\begin{equation}
\rho(y,t) = B_{n}^2 y^{2\alpha } e 
^{\frac{F}{\alpha} y^{2\alpha }}\left(\frac{d^{n}}{dy^{n}}\left[ 
y^{n-\alpha}e^{\frac{-F}{\alpha } y^{2\alpha}}\right]\right)^{2} e^{-\frac{\lambda t^{\alpha}}{2}}.
\end{equation}
Due to dissipation, the eigenfunctions are normalized at the initial time $t=0$ by the relation:
\begin{equation}
\int_0 ^\infty \rho(y,0)dy=1,
\end{equation}
which fixes the normalization constant $B_{n}$ at all times. To check the calculations, we plotted in Figure (\ref{fig1}) the time evolution of the probability density for the ground state $n=0$  and the first excited state $n=1$ with parameter set $\lambda=0$ and $\alpha=1$. This assumes to represent the traditional quantum harmonic oscillator with no dissipation. It is clear that we have demonstrated here the well-known behavior for this situation. In Figure (\ref{fig2}), we plotted the same situation $\alpha=1$ but with dissipation $\lambda=0.5$. It is found there is a decrease in the probability density as the system evolves with time. This is due to dissipation as expected. However, the behavior of the probability distribution decrease found in this work is such that not only the areas under the curves  decrease as time evolves, but also the peaks get lowered gradually. In comparison with the results found in \cite{serhanQuantizationDampedHarmonic2018}, their distributions decreases but keeping same peaks. Figure (\ref{fig3}) shows the behavior of the initial probability density distributions for the first three states $n=0,1,2,3$ and the parameter $F=1$, evaluated at different values of the fractional order parameter $\alpha$. It can be concluded that there is a trend toward a gradual ordering in these distributions with $\alpha$. This could suggest to use the conformable  model to describe or model nonlinear phenomena accompanied the original Bateman system.  

 \clearpage

\section{Conclusion}
\label{sec3}
In this work, the Bateman’s Lagrangian for the damped harmonic oscillator system is formulated using the conformable  derivative concept, from which we then obtained the corresponding conformable  Euler-Lagrange equations of motion and  conformable Hamiltonian. The system is then canonically quantized, which led to the appropriate  conformable Schrodinger equation. Using some transformations and the extended conformable  Nikiforov-Uvarov method, we were able to solve the equation and obtain the energy eigenvalues and eigenfunctions as a function of the $\alpha$ order. The energy eigenvalues were found to be real and equally spaced and the corresponding traditional damped harmonic oscillator was recovered by setting $\alpha=1$.  We derived the continuity equation to define the probability density and probability current. The probability density evolution as well as its dependence on the $\alpha$ order was plotted and analyzed for various situations. We observed a gradual ordering  in these distribution for $0.8\leq \alpha \leq 1$.

\bibliography{ref} 

\begin{thebibliography}{10}
\providecommand{\url}[1]{#1}
\csname url@samestyle\endcsname
\providecommand{\newblock}{\relax}
\providecommand{\bibinfo}[2]{#2}
\providecommand{\BIBentrySTDinterwordspacing}{\spaceskip=0pt\relax}
\providecommand{\BIBentryALTinterwordstretchfactor}{4}
\providecommand{\BIBentryALTinterwordspacing}{\spaceskip=\fontdimen2\font plus
\BIBentryALTinterwordstretchfactor\fontdimen3\font minus
  \fontdimen4\font\relax}
\providecommand{\BIBforeignlanguage}[2]{{%
\expandafter\ifx\csname l@#1\endcsname\relax
\typeout{** WARNING: IEEEtran.bst: No hyphenation pattern has been}%
\typeout{** loaded for the language `#1'. Using the pattern for}%
\typeout{** the default language instead.}%
\else
\language=\csname l@#1\endcsname
\fi
#2}}
\providecommand{\BIBdecl}{\relax}
\BIBdecl

\bibitem{batemanDissipativeSystemsRelated1931}
H.~Bateman, ``\BIBforeignlanguage{en}{On {{Dissipative Systems}} and {{Related
  Variational Principles}}},'' \emph{\BIBforeignlanguage{en}{Physical Review}},
  vol.~38, no.~4, pp. 815--819, Aug. 1931.

\bibitem{caldirolaForzeNonConservative1941}
P.~Caldirola, ``\BIBforeignlanguage{it}{{Forze non conservative nella meccanica
  quantistica}},'' \emph{\BIBforeignlanguage{it}{Il Nuovo Cimento}}, vol.~18,
  no.~9, pp. 393--400, Nov. 1941.

\bibitem{kanaiQuantizationDissipativeSystems1948}
E.~Kanai, ``\BIBforeignlanguage{en}{On the {{Quantization}} of the
  {{Dissipative Systems}}},'' \emph{\BIBforeignlanguage{en}{Progress of
  Theoretical Physics}}, vol.~3, no.~4, pp. 440--442, Dec. 1948.

\bibitem{jainPathIntegralsWavepacket2007}
D.~Jain, A.~Das, and S.~Kar, ``\BIBforeignlanguage{en}{Path integrals and
  wavepacket evolution for damped mechanical systems},''
  \emph{\BIBforeignlanguage{en}{American Journal of Physics}}, vol.~75, no.~3,
  pp. 259--267, Mar. 2007.

\bibitem{baldiottiQuantizationDampedHarmonic2011}
M.~Baldiotti, R.~Fresneda, and D.~Gitman,
  ``\BIBforeignlanguage{en}{Quantization of the damped harmonic oscillator
  revisited},'' \emph{\BIBforeignlanguage{en}{Physics Letters A}}, vol. 375,
  no.~15, pp. 1630--1636, Apr. 2011.

\bibitem{markusQuantumParticleMotion2016}
B.~G. M{\'a}rkus and F.~M{\'a}rkus, ``\BIBforeignlanguage{en}{Quantum particle
  motion in absorbing harmonic trap},'' \emph{\BIBforeignlanguage{en}{Indian
  Journal of Physics}}, vol.~90, no.~4, pp. 441--446, Apr. 2016.

\bibitem{serhanQuantizationDampedHarmonic2018}
M.~Serhan, M.~Abusini, A.~{Al-Jamel}, H.~{El-Nasser}, and E.~M. Rabei,
  ``\BIBforeignlanguage{en}{Quantization of the damped harmonic oscillator},''
  \emph{\BIBforeignlanguage{en}{Journal of Mathematical Physics}}, vol.~59,
  no.~8, p. 082105, Aug. 2018.

\bibitem{takahashiQuantizationMassiveBateman2018}
K.~Takahashi, ``On the quantization of the massive {{Bateman}} system,''
  \emph{Journal of Mathematical Physics}, vol.~59, no.~7, p. 072108, Jul. 2018.

\bibitem{oldham1974fractional}
K.~Oldham and J.~Spanier, ``The fractional calculus, academic press, new
  york,'' \emph{The fractional calculus. Academic Press, New York.}, 1974.

\bibitem{miller1993introduction}
K.~Miller and B.~Ross, ``An introduction to the fractional integrals and
  derivatives-theory and applications. {{John}} willey \& sons,'' \emph{Inc.,
  New York}, 1993.

\bibitem{kilbas2006theory}
A.~A. Kilbas, H.~M. Srivastava, and J.~J. Trujillo, \emph{Theory and
  Applications of Fractional Differential Equations}.\hskip 1em plus 0.5em
  minus 0.4em\relax {elsevier}, 2006, vol. 204.

\bibitem{klimek2002lagrangean}
M.~Klimek, ``Lagrangean and {{Hamiltonian}} fractional sequential mechanics,''
  \emph{Czechoslovak Journal of Physics}, vol.~52, no.~11, pp. 1247--1253,
  2002.

\bibitem{agrawal2002formulation}
O.~P. Agrawal, ``Formulation of {{Euler}}\textendash{{Lagrange}} equations for
  fractional variational problems,'' \emph{Journal of Mathematical Analysis and
  Applications}, vol. 272, no.~1, pp. 368--379, 2002.

\bibitem{rieweNonconservativeLagrangianHamiltonian1996}
F.~Riewe, ``\BIBforeignlanguage{en}{Nonconservative {{Lagrangian}} and
  {{Hamiltonian}} mechanics},'' \emph{\BIBforeignlanguage{en}{Physical Review
  E}}, vol.~53, no.~2, pp. 1890--1899, Feb. 1996.

\bibitem{rieweMechanicsFractionalDerivatives1997}
------, ``\BIBforeignlanguage{en}{Mechanics with fractional derivatives},''
  \emph{\BIBforeignlanguage{en}{Physical Review E}}, vol.~55, no.~3, pp.
  3581--3592, Mar. 1997.

\bibitem{rabeiQuantizationFractionalSystems2010}
E.~M. Rabei, S.~I. Muslih, and D.~Baleanu,
  ``\BIBforeignlanguage{en}{Quantization of fractional systems using {{WKB}}
  approximation},'' \emph{\BIBforeignlanguage{en}{Communications in Nonlinear
  Science and Numerical Simulation}}, vol.~15, no.~4, pp. 807--811, Apr. 2010.

\bibitem{rabeiQuantizationBrownianMotion2006}
E.~M. Rabei, A.-W. Ajlouni, and H.~B. Ghassib,
  ``\BIBforeignlanguage{en}{Quantization of {{Brownian Motion}}},''
  \emph{\BIBforeignlanguage{en}{International Journal of Theoretical Physics}},
  vol.~45, no.~9, pp. 1613--1623, Nov. 2006.

\bibitem{rabei2009fractional}
E.~M. Rabei, I.~M. Altarazi, S.~I. Muslih, and D.~Baleanu, ``Fractional {{WKB}}
  approximation,'' \emph{Nonlinear Dynamics}, vol.~57, no. 1-2, pp. 171--175,
  2009.

\bibitem{rabei2006quantization}
E.~M. Rabei, A.-W. Ajlouni, and H.~B. Ghassib, ``Quantization of brownian
  motion,'' \emph{International Journal of theoretical physics}, vol.~45,
  no.~9, pp. 1613--1623, 2006.

\bibitem{khalilNewDefinitionFractional2014}
R.~Khalil, M.~Al~Horani, A.~Yousef, and M.~Sababheh,
  ``\BIBforeignlanguage{en}{A new definition of fractional derivative},''
  \emph{\BIBforeignlanguage{en}{Journal of Computational and Applied
  Mathematics}}, vol. 264, pp. 65--70, Jul. 2014.

\bibitem{khalil2014conformable}
R.~Khalil and M.~{Abu-Hammad}, ``Conformable fractional heat differential
  equation,'' \emph{International Journal of Pure and Applied Mathematics},
  vol.~94, pp. 215--217, 2014.

\bibitem{lazo2016variational}
M.~J. Lazo and D.~F. Torres, ``Variational calculus with conformable fractional
  derivatives,'' \emph{IEEE/CAA Journal of Automatica Sinica}, vol.~4, no.~2,
  pp. 340--352, 2016.

\bibitem{chung2020effect}
W.~S. Chung, S.~Zare, H.~Hassanabadi, and E.~Maghsoodi, ``The effect of
  fractional calculus on the formation of quantum-mechanical operators,''
  \emph{Mathematical Methods in the Applied Sciences}, 2020.

\bibitem{al2019search}
A.~{Al-Jamel}, ``The search for fractional order in heavy quarkonia spectra,''
  \emph{International Journal of Modern Physics A}, vol.~34, no.~10, p.
  1950054, 2019.

\bibitem{karayerConformableFractionalNikiforov2016}
H.~Karayer, D.~Demirhan, and F.~B{\"u}y{\"u}kk{\i}l{\i}{\c c}, ``Conformable
  {{Fractional Nikiforov}}\textemdash{{Uvarov Method}},'' \emph{Communications
  in Theoretical Physics}, vol.~66, no.~1, pp. 12--18, Jul. 2016.

\bibitem{karayerSolutionsLocalFractional2019}
H.~Karayer, D.~Demirhan, and F.~Buyukkilic, ``\BIBforeignlanguage{en}{Solutions
  of local fractional sine-{{Gordon}} equations},''
  \emph{\BIBforeignlanguage{en}{Waves in Random and Complex Media}}, vol.~29,
  no.~2, pp. 227--235, Apr. 2019.

\bibitem{chungFractionalNewtonMechanics2015}
W.~S. Chung, ``\BIBforeignlanguage{en}{Fractional {{Newton}} mechanics with
  conformable fractional derivative},'' \emph{\BIBforeignlanguage{en}{Journal
  of Computational and Applied Mathematics}}, vol. 290, pp. 150--158, Dec.
  2015.

\bibitem{rabei2018quantization}
E.~M. Rabei and M.~Al~Horani, ``Quantization of fractional singular
  {{Lagrangian}} systems using {{WKB}} approximation,'' \emph{International
  Journal of Modern Physics A}, vol.~33, no.~36, p. 1850222, 2018.

\bibitem{https://doi.org/10.1002/mma.7963}
\BIBentryALTinterwordspacing
M.~Al-Masaeed, E.~M. Rabei, and A.~Al-Jamel, ``Extension of the variational
  method to conformable quantum mechanics,'' \emph{Mathematical Methods in the
  Applied Sciences}, vol.~45, no.~5, pp. 2910--2920, 2022. [Online]. Available:
  \url{https://onlinelibrary.wiley.com/doi/abs/10.1002/mma.7963}
\BIBentrySTDinterwordspacing

\bibitem{al2021extension}
M.~Al-Masaeed, E.~M. Rabei, A.~Al-Jamel, and D.~Baleanu, ``Extension of
  perturbation theory to quantum systems with conformable derivative,''
  \emph{Modern Physics Letters A}, p. 2150228, 2021.

\bibitem{doi:10.1142/S0217732322501449}
\BIBentryALTinterwordspacing
M.~Al-Masaeed, E.~M. Rabei, and A.~Al-Jamel, ``Wkb approximation with
  conformable operator,'' \emph{Modern Physics Letters A}, vol.~37, no.~22, p.
  2250144, 2022. [Online]. Available:
  \url{https://doi.org/10.1142/S0217732322501449}
\BIBentrySTDinterwordspacing

\bibitem{al2021quantization}
M.~Al-Masaeed, E.~M. Rabei, A.~Al-Jamel, and D.~Baleanu, ``Quantization of
  fractional harmonic oscillator using creation and annihilation operators,''
  \emph{Open Physics}, vol.~19, no.~1, pp. 395--401, 2021.

\bibitem{al-jamel_effect_2022}
\BIBentryALTinterwordspacing
A.~Al-Jamel, M.~Al-Masaeed, E.~Rabei, and D.~Baleanu,
  ``\BIBforeignlanguage{en}{The effect of deformation of special relativity by
  conformable derivative},'' \emph{\BIBforeignlanguage{en}{Revista Mexicana de
  Física}}, vol.~68, no. 5 Sep-Oct, pp. 050\,705 1--9, Aug. 2022, number: 5
  Sep-Oct. [Online]. Available:
  \url{https://rmf.smf.mx/ojs/index.php/rmf/article/view/5877}
\BIBentrySTDinterwordspacing

\bibitem{https://doi.org/10.48550/arxiv.2112.01322}
\BIBentryALTinterwordspacing
E.~M. Rabei, A.~Al-Jamel, and M.~Al-Masaeed, ``The solution of conformable
  laguerre differential equation using conformable laplace transform,'' 2021.
  [Online]. Available: \url{https://arxiv.org/abs/2112.01322}
\BIBentrySTDinterwordspacing

\bibitem{https://doi.org/10.48550/arxiv.2203.11615}
\BIBentryALTinterwordspacing
E.~M. Rabei, M.~Al-Masaeed, and A.~Al-Jamel, ``Solution of the conformable
  angular equation of the schrodinger equation,'' 2022. [Online]. Available:
  \url{https://arxiv.org/abs/2203.11615}
\BIBentrySTDinterwordspacing

\bibitem{https://doi.org/10.48550/arxiv.2209.02699}
\BIBentryALTinterwordspacing
M.~Al-Masaeed, E.~M. Rabei, and A.~Al-Jamel, ``Analytic study of conformable
  schrodinger equation with hydrogen atom,'' 2022. [Online]. Available:
  \url{https://arxiv.org/abs/2209.02699}
\BIBentrySTDinterwordspacing

\bibitem{karayerExtensionNikiforovUvarovMethod2015}
H.~Karayer, D.~Demirhan, and F.~B{\"u}y{\"u}kk{\i}l{\i}{\c c},
  ``\BIBforeignlanguage{en}{Extension of {{Nikiforov}}-{{Uvarov}} method for
  the solution of {{Heun}} equation},'' \emph{\BIBforeignlanguage{en}{Journal
  of Mathematical Physics}}, vol.~56, no.~6, p. 063504, Jun. 2015.

\bibitem{al-jamelSaturationHeavyQuarkonia2018b}
A.~{Al-Jamel}, ``\BIBforeignlanguage{en}{Saturation in heavy quarkonia spectra
  with energy-dependent confining potential in {{{\emph{N}}}} -dimensional
  space},'' \emph{\BIBforeignlanguage{en}{Modern Physics Letters A}}, vol.~33,
  no.~32, p. 1850185, Oct. 2018.

\bibitem{karayerSolutionSchrodingerEquation2018}
H.~Karayer, D.~Demirhan, and F.~B{\"u}y{\"u}kk{\i}l{\i}{\c c},
  ``\BIBforeignlanguage{en}{Solution of {{Schr\"odinger}} equation for two
  different potentials using extended {{Nikiforov}}-{{Uvarov}} method and
  polynomial solutions of biconfluent {{Heun}} equation},''
  \emph{\BIBforeignlanguage{en}{Journal of Mathematical Physics}}, vol.~59,
  no.~5, p. 053501, May 2018.

\bibitem{mozaffari2018conformable}
F.~Mozaffari, H.~Hassanabadi, H.~Sobhani, and W.~Chung, ``On the conformable
  fractional quantum mechanics,'' \emph{Journal of the Korean Physical
  Society}, vol.~72, no.~9, pp. 980--986, 2018.

\end{thebibliography}
\bibliographystyle{IEEEtran}
\end{document}